\documentclass[twocolumn,showpacs,preprintnumbers,amsmath,amssymb,a4paper,prl]{revtex4}
\usepackage{graphicx}
\usepackage{amssymb, amsmath,amssymb,amsfonts}
\usepackage{graphicx}
\usepackage{dcolumn}
\usepackage{bm}
\usepackage{color}
\begin{document}
\title{Stochastic Turing Patterns on a Network}
\author{Malbor Asslani}
\affiliation{Dipartimento di  Scienza e Alta Tecnologia, Universit\`{a} degli Studi dell'Insubria, via Valleggio 11, 22100 Como, Italy}
\author{Francesca Di Patti}
\affiliation{Dipartimento di Energetica ``Sergio Stecco'', Universit\`{a} degli Studi di Firenze, via S. Marta 3, 50139 Firenze, Italy and INFN, Sezione di Firenze}
\author{Duccio Fanelli}
\affiliation{Dipartimento di Energetica ``Sergio Stecco'', Universit\`{a} degli Studi di Firenze, via S. Marta 3, 50139 Firenze, Italy and INFN, Sezione di Firenze}
\begin{abstract}
The process of stochastic Turing instability on a network is discussed for a specific case study, the stochastic Brusselator model. 
The system is shown to spontaneously differentiate into activator-rich and activator-poor nodes, outside the region of parameters classically deputed to the deterministic Turing instability. This phenomenon, as revealed by direct stochastic simulations, is explained analytically, and eventually traced back to the finite size corrections stemming from the inherent graininess of the scrutinized medium. 
\end{abstract}
\maketitle
Pattern formation is a rich and fascinating field of investigation which extends over distinct realms of applications, ideally embracing chemistry, biology, and physics. Complex and extremely beautiful patterns can in fact spontaneously emerge in spatially extended reaction-diffusion systems, as follows a linear instability mechanism, first described by Alan Turing in a seminal paper \cite{turing} published in 1952. Turing patterns are indeed widespread in nature: examples include schemes of autocatalytic reactions with inhibition \cite{prigogine,castets,quyang}, the process of biological morphogenesis \cite{meinhardt, harris, maini, bhat, miura} and the dynamics of ecosystems \cite{mimura,maron,baurmann,riet}. The Turing instability paradigm  classically relies on mean field, deterministic scenarios. As opposed to the usual continuum picture, the intimate discreteness of any individual based stochastic models results in finite size corrections to the approximated mean-field dynamics. Under specific conditions, microscopic disturbances are enhanced as follows a resonance mechanism and yield organized spatio-temporal patterns \cite{mckanePRL, goldenfeld, biancalani, woolley}. More specifically, the measured concentration which reflects the distribution of the interacting entities (e.g. chemical species, biomolecules) can display spatially patched profile, collective phenomena which testify on a surprising degree of macroscopic order, as mediated by the stochastic component of the dynamics. Stochastic Turing patterns \cite{biancalani}, or quasi Turing patterns \cite{goldenfeld}, are found to occur in individual based systems, that cannot undergo Turing instability according to the deterministic reaction-diffusion picture. Interestingly, the region of parameters for which stochastic patterns are developed is usually larger than for conventional Turing patterns, a general observation that has been made quantitative for a selection of prototypical case studies.   

Recently, Nakao and Mikhailov \cite{nakao} studied the Turing patterns formation on large random networks, an important direction of investigation presumably relevant  in e.g. the early stage of the morphogenesis process, since morphogens are known to diffuse on the network structure of inter-cellular connections. Already in 1971 Othmer and Scriver \cite{othmer} investigated the Turing instability in network-organized system and developed the needed mathematical machineries. Their studies were however limited to regular lattice or small networks. By extending the analysis to complex heterogeneous network Nakao and Mikhailov \cite{nakao} opened up the perspective for novel applications of the Turing idea to the broad field of theoretical biology \cite{vespignani}. Applications can be also foreseen in other disciplines were network science proves crucial. Among others, social studies, in which nodes and links are respectively associated to humans and their mutual interactions, and the analysis of epidemics spreading, which reflects the topological structure of the underlying mobility networks. 

Starting from this setting, we here propose a generalization of the work \cite{nakao}, beyond the deterministic scenario, by explicitly including the role of demographic, finite size fluctuations. In doing so, we will demonstrate in this Letter  that {\it Stochastic Turing patterns} set in, outside the region of parameters deputed to spatial order, as predicted within the classical theory based on deterministic reaction-diffusion schemes. 

To this end, we will consider a stochastic version of the celebrated Brusselator model \cite{prigogine} which will be placed on top of an heterogeneous network of $\Omega$ nodes. More concretely, two species, respectively  $X_i$ and $Y_i$ are assigned to the generic node $i$, and therein react according to the following chemical reactions \cite{biancalani}:
\begin{eqnarray}
A+E_i       &\stackrel{a}{\longrightarrow}  &  A+X_i \nonumber \\
X_i+B       &\stackrel{b}{\longrightarrow}  &  Y_i+B  \nonumber \\
2X_i+Y_i  &\stackrel{c}{\longrightarrow}  &  3X_i \nonumber \\
X_i           &\stackrel{d}{\longrightarrow } &  E_i    \label{chemical} \quad .
\end{eqnarray}
The symbol $E_i$ stands for an empty case and formally amounts to imposing a finite carrying capacity in each node of the network. In other words, we assume that each node can host a maximum number $N$ of molecules (or agents), including the vacancies \footnote{For a discussion of the role played by finite carrying capacity we refer to \cite{mckanePRL,mckanefanelli, lugo, biancalani2}. The forthcoming analysis can be repeated by relaxing such an assumption, and yielding qualitative equivalent results.}. Let us denote by $n_i$ and $m_i$ the total number of elements belonging to species $X$ and $Y$ in the node $i$. The corresponding  number of empties total hence in $N-n_i-m_i$. The parameters $a$, $b$, $c$ and $d$ in Eqs. (\ref{chemical}) are the reaction rates, while the species $A$ and $B$ are enzymatic activators whose concentrations  are supposed to remain constant during the dynamics. In addition to the above activator-inhibitor rules, we assume that the molecules can migrate between neighbour nodes as dictated by the following reactions: 
\begin{eqnarray}
X_i+E_j  &\stackrel{\mu}{\longrightarrow}     &  E_i+X_j \nonumber \\
Y_i+E_j  &\stackrel{\delta}{\longrightarrow}  &  E_i+Y_j \label{eq:chem2}
\end{eqnarray}
where $\mu$ and $\delta$ are the diffusion coefficients characteristic of the two species, and the subscript $j$ denotes the generic node connected to $i$, via the network structure. Similar equations governs the diffusion from node $j$ towards node $i$. To complete the notation we introduce the $\Omega$-dimensional vectors $\mathbf{n}=(n_1,...,n_i,...,n_\Omega)$ 
and $\mathbf{m}=(m_1,...,m_i,...,m_\Omega)$  that unequivocally identify the state of the system. The process here 
imagined is intrinsically stochastic. Under the Markov hypothesis, the probability $P(\mathbf{n},\mathbf{m},t)$ of seeing the system at time $t$ in state ($\mathbf{n}$,$\mathbf{m}$) obeys to a master equation that can be cast in the compact form:
\begin{small}
\begin{eqnarray}
\frac{\partial}{\partial t}P(\mathbf{n},\mathbf{m},t)&=&\sum_{i=1}^\Omega\Big\{(\epsilon_{n_i}^{-} -1)T(n_{i}+1, m_{i}|n_{i},m_{i}) \nonumber\\
&+&(\epsilon_{n_i}^{+} -1)T(n_{i}-1, m_{i}|n_{i},m_{i})\nonumber\\
&+&(\epsilon_{n_i}^{-}\epsilon_{m_i}^{+}-1)T(n_{i}+1, m_{i}-1|n_{i},m_{i}) \nonumber\\
&+&(\epsilon_{n_i}^{+}\epsilon_{m_i}^{-} -1)T(n_{i}-1, m_{i}+1|n_{i},m_{i})\nonumber\\
&+&\sum_{j=1}^\Omega W_{i,j}\Big[(\epsilon_{n_i}^{+}\epsilon_{n_j}^{-} -1)T(n_{i}-1, n_{j}+1|n_{i},n_{j}) \nonumber\\
&+&(\epsilon_{n_j}^{+}\epsilon_{n_i}^{-} -1)T(n_{j}-1, n_{i}+1|n_{i},n_{j})\nonumber\\
&+&(\epsilon_{m_i}^{+}\epsilon_{m_j}^{-} -1)T(m_{i}-1, m_{j}+1|m_{i},m_{j}) \nonumber\\
&+&(\epsilon_{m_j}^{+}\epsilon_{m_i}^{-} -1)T(m_{j}-1, m_{i}+1 |m_{i},m_{j})\big]\Big\}  \nonumber \\
&\times& P(\mathbf{n}, \mathbf{m}, t)\label{eq:master}
\end{eqnarray}
\end{small}
where use has been made of the step operators $\epsilon_{n_i}^\pm f(...,n_i,..., \mathbf{m})=f(...,n_i \pm 1,..., \mathbf{m})$ and
$\epsilon_{m_i}^\pm f(\mathbf{n}, ...,m_i,...)= f(\mathbf{n}, ...,m_i \pm 1,...)$, $f(\cdot,\cdot)$ being any generic function of the state variables. The $\Omega \times \Omega$ integers $W_{i,j}$ represent the entries of the symmetric adjacency matrix $\mathbf{W}$, which characterizes the topology of the network. $W_{i,j}$ is equal to $1$ if nodes $i$ and $j$ are connected, and $0$ otherwise.  The transition rates $T(\mathbf{n}',\mathbf{m}'|\mathbf{n},\mathbf{m})$ link the initial state $(\mathbf{n},\mathbf{m})$ to another state $(\mathbf{n}',\mathbf{m}')$, compatible with the former, and are given by 
\begin{small}
\begin{eqnarray*}
T(n_{i}+1, m_{i}|n_{i},m_{i})&=&\frac{a}{\Omega}\frac{N-n_i-m_i}{N}\\
T(n_{i}-1, m_{i}|n_{i},m_{i})&=&\frac{d}{\Omega}\frac{n_i}{N}\\
T(n_{i}+1, m_{i}-1|n_{i},m_{i})&=&\frac{c}{\Omega}\frac{n_i^2m_i}{N^3}\\
T(n_{i}-1, m_{i}+1|n_{i},m_{i})&=&\frac{b}{\Omega}\frac{n_i}{N}\\
T(n_{i}-1, n_{j}+1|n_{i},n_{j})&=&\frac{\mu}{\Omega}\frac{n_i}{N}\frac{N-n_j-m_j}{N}\left(\frac{1}{k_i}+\frac{1}{k_j}\right)\\
T(m_{i}-1, m_{j}+1|m_{i},m_{j})&=&\frac{\delta}{\Omega}\frac{m_i}{N}\frac{N-n_j-m_j}{N}\left(\frac{1}{k_i}+\frac{1}{k_j}\right).
\end{eqnarray*}\end{small}
where $k_i=\sum_{j=1}^\Omega W_{i,j}$ is the degree of the $i-$th node. The factor $1/k_i+1/k_j$ takes into account the order of selection of species in chemical reactions (\ref{eq:chem2}). 

The master equation (\ref{eq:master}) is exact although difficult to handle. To progress in the analysis it is customary to resort to approximated perturbation methods. In the weak noise approximation, one can put forward the  van Kampen ansatz \cite{vankampen, gardiner}  which, in this context, amounts to imposing $n_i/N=\phi_i+\xi_{1i} / \sqrt{N}$ and $m_i=\psi_i+\xi_{2i} / \sqrt{N}$. $\phi_i$ and $\psi_i$ are the mean field concentrations respectively associated to the interacting species $X$ and $Y$. $\xi_{1i}$ and $\xi_{2i}$ are stochastic fluctuations that originate from finite size corrections, normalized by the scaling factor $1/ \sqrt{N}$, as dictated by the central limit theorem \cite{vankampen}. For moderately large system sizes $N$,  the $1/\sqrt{N}$ factor is small and paves the way to a straightforward perturbative calculation, generally referred to in the literature as to the van Kampen system size expansion. At the leading order of the perturbative analysis, the mean field equations for the deterministic variables are recovered and, for the specific problem here investigated, read: 
 \begin{multline}
\frac{d}{d\tau}\phi_i= f(\phi_i, \psi_i)+2\mu \left[ 
 \sum_{j=1}^\Omega\Delta_{ij}  \phi_j +\phi_i \sum_{j=1}^\Omega\Delta_{ij} \psi_j \right.
\\ \left.-\psi_i \sum_{j=1}^\Omega\Delta_{ij}  \phi_j  \right] \\
\frac{d}{d\tau} \psi_i=g(\phi_i, \psi_i)+2\delta  \left[ 
\sum_{j=1}^\Omega\Delta_{ij}\psi_j + \phi_i \sum_{j=1}^\Omega\Delta_{ij} \phi_j \right.
\\\left. -\phi_i \sum_{j=1}^\Omega\Delta_{ij} \psi_j \right] \label{eq:mf}
\end{multline}
where, generalizing the heuristic derivation of \cite{nakao}, we have introduced the discrete Laplacian $\Delta_{ij}=\tilde{W}_{ij}-\tilde{k}_i\delta_{ij}$ 
with  $\tilde{k}_i=\sum_{j=1}^\Omega\tilde{W}_{ij}$ and $\tilde{W}_{ij} = (1/k_i+1/k_j) W_{i,j}$. The reaction terms are respectively $f=-(b+d)\phi_i+c\phi_i^2\psi_i+a(1-\phi_i-\psi_i)$ and $g=b\phi_i-c\phi_i^2\psi_i$.  $\tau$ is the rescaled time $t/(N \Omega)$.  Cross diffusion terms appear in the obtained deterministic equations, because of the finite carrying capacity, imposed at the level of the single node \cite{mckanefanelli}. By relaxing such an assumption \cite{biancalani2}, conventional diffusion operators are instead recovered. Similarly, the finite carrying capacity assumption reflects in the reaction contribution $a(1-\phi_i-\psi_i)$ that replaces the usual constant term $a$ in the standard Brusselator equations \cite{biancalani}. Although interesting per se, this modification does not play any substantial role in the forthcoming development: equivalent conclusions can be drawn when working in the diluted setting, i.e. away from jamming or crowding conditions that inspire the physically sound request for a limited capacity to be explicitly accommodated on each individual node.   
%%%%%%%%%%%%%%%%%%%%%%%%%%%%%%%%%%%%%%%%%%%%%%
\begin{figure}[tb]
\begin{center}
\includegraphics[scale=0.3]{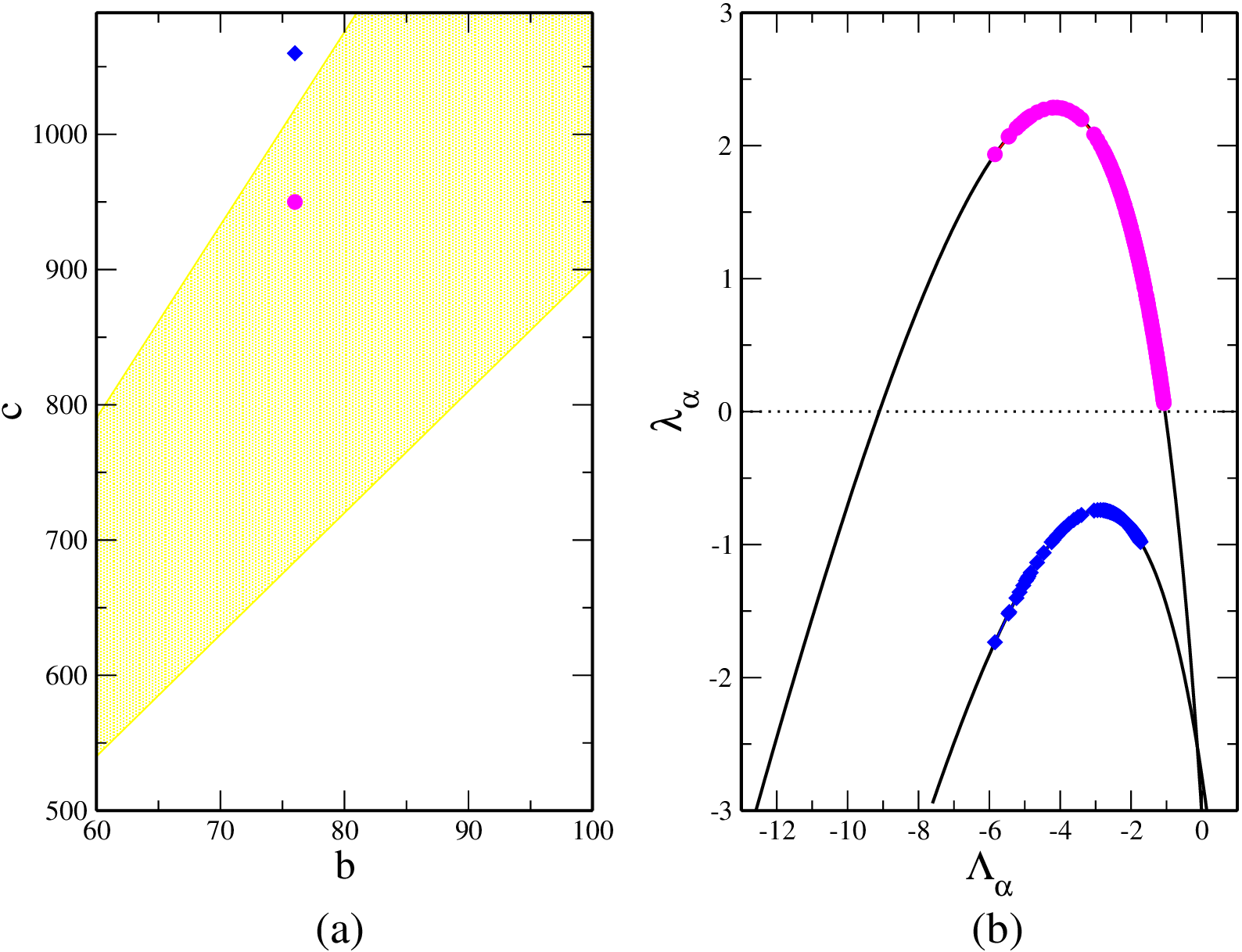}
\end{center}
\caption{The darkened region (yellow on line)  in panel (a) delineates the Turing instability domain in the $(b,c)$ plane for the Brusselator model with $a=d=1$, $\mu=1$ and $\delta=15$. The (magenta on line) point belongs to the Turing instability region and corresponds to $b=76$ and  $c=950$. The (blue online) diamond falls outside the region of Turing order and is positioned at $(76,1060)$. In panel (b) the dispersion relation (\ref{eq:dispersion}) is plotted as a function of both the discrete eigenvalues of the network Laplacian (symbols) and their real analogues $-k^2$ (solid line). Circles (magenta online) refer to $(b,c)=(76,950)$, while diamonds (bue online) to $(b,c)=(76,1060)$. In the analysis we assumed a scale-free network made of $\Omega=200$ nodes and mean degree $\langle  k \rangle=20$. The network has been generated according to the Barab\'{a}si-Albert algorithm \cite{barabasi}.}\label{fig:dispersion}
\end{figure}
%%%%%%%%%%%%%%%%%%%%%%%%%%%%%%%%%%%%%%%%%%%%%%
%%%%%%%%%%%%%%%%%%%%%%%%%%%%%%%%%%%%%%%%%%%%%%
\begin{figure}[tb]
\begin{center}
\includegraphics[scale=0.35]{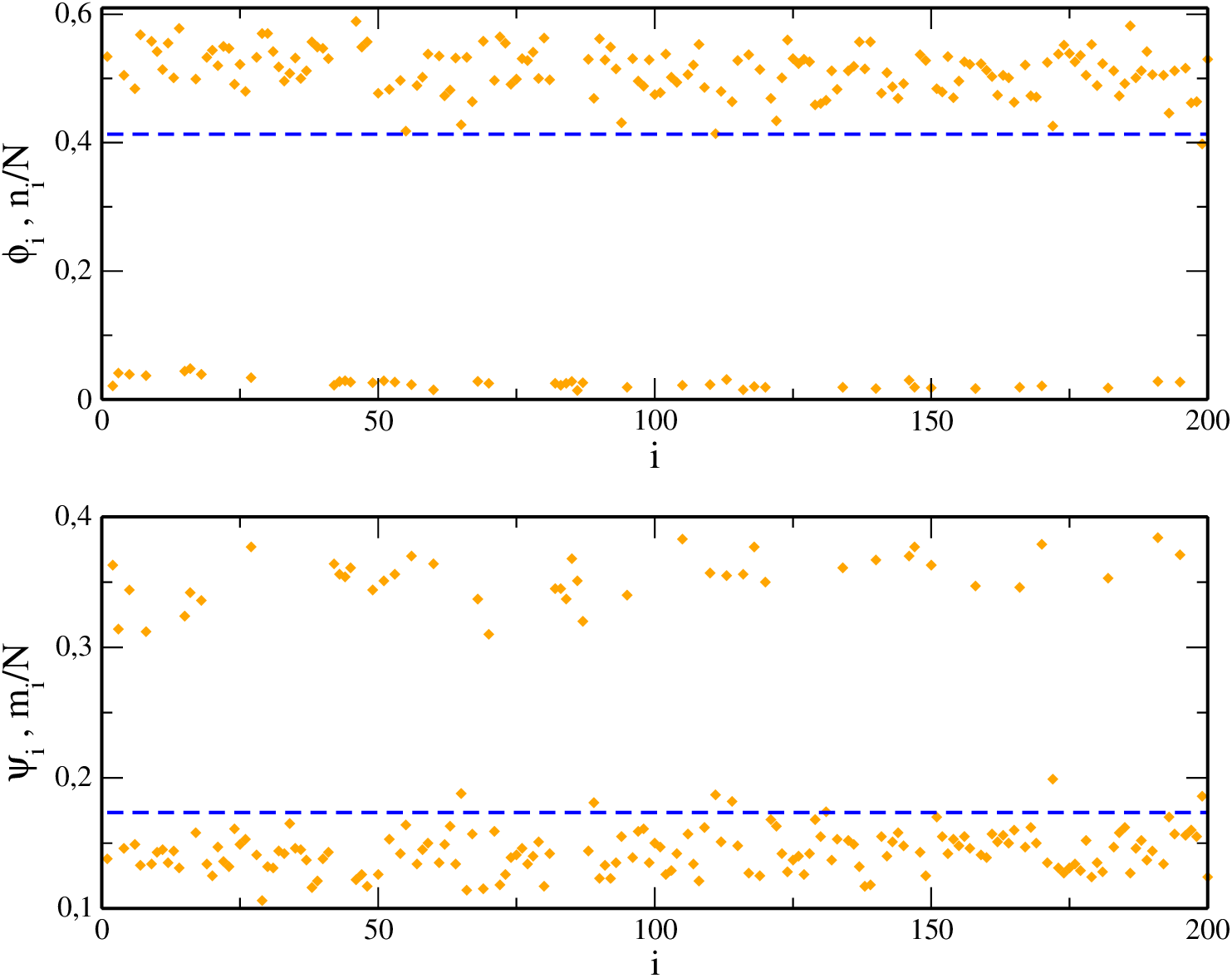}
\end{center}
\caption{Simulations of the stochastic chemical model (\ref{chemical})-(\ref{eq:chem2}) outside the region of Turing order, $a=d=1$, $b=76$, $c=1060$, $\mu=1$, $\delta=15$. Here $N=1000$. The late time concentrations per node $n_i/N$ (resp. $m_i/N$) are plotted in the upper panel (resp. lower) panel, as a function of the node index $i$. The (orange online) diamonds are obtained from one realization of the stochastic Gillespie algorithm \cite{gillespie}. The network is generated as described in the caption of Fig. \ref{fig:dispersion}. The stochastic dynamics yields  the emergence of two distinct activator-rich and activator-poor groups, while the deterministic dynamics is attracted towards the stable (and trivial) homogeneous fixed point, dashed (blue online) horizontal line. \label{fig:stocSim}}
\end{figure}
%%%%%%%%%%%%%%%%%%%%%%%%%%%%%%%%%%%%%%%%%%%%%%

To look for mean-field Turing instability, one needs to introduce a small perturbation to the homogeneous equilibrium point $(\phi^*,\psi^*)=((a+\sqrt{a^2-4ab(a+d)/c})/2/(a+d),b/c/\phi^*)$ of the deterministic system (\ref{eq:mf}) and carry out a linear stability analysis. In formulae, $(\phi_i,\psi_i)=(\phi^*+\delta \phi_i,\psi^*+\delta \psi_i)$.  Following \cite{nakao}, and to exploit the linearity of the resulting equations for the perturbation amounts \footnote{It is straightforward to show that the perturbations  obey to $\delta \dot{\phi}_i=f_{\phi}\delta \phi_i+f_{\psi}\delta \psi_i $ $+\mu\left[(1-\psi^*)\sum_{j=1}\Delta_{ij}\delta\phi_j \right. $ $ \left. +\phi^*\sum_{j=1}\Delta_{ij}\delta \psi_j\right]$ and $\delta \dot{\psi}_i=g_{\phi}\delta \phi_i+g_{\psi}\delta \psi_i +\delta\left[\left(1-\phi^*\right)\sum_{j=1}\Delta_{ij}\delta\psi_j   +\psi^*\sum_{j=1}\Delta_{ij}\delta \phi_j\right]$, under the linear approximation.}, we find it convenient to expand $\delta\phi_i$ and $\delta\phi_i$ as:
\begin{equation}\label{eq:modes}
\delta\phi_i =\sum_{\alpha=1}^\Omega c_\alpha e^{\lambda_\alpha\tau} v_i^{(\alpha)} \qquad
\delta\psi_i=\sum_{\alpha=1}^\Omega c_\alpha\beta_\alpha e^{\lambda_\alpha\tau} v_i^{(\alpha)}
\end{equation}
where $\mathbf{v}^{(\alpha)}=(v_1^{(\alpha)}, \ldots , v_{\Omega}^{(\alpha)})$ stand for the eigenvectors of the 
Laplacian operator corresponding to the eigenvalue $\Lambda_{\alpha}$ \footnote{The Laplacian  operator $\Delta_{i,j}$ is defined by the  real and symmetric matrix $\Delta_{ij}=\tilde{W}_{ij}-\tilde{k}_i\delta_{ij}$ where $k_i$ is the degree of node $i$. Eigenvectors $\mathbf{v}^{(\alpha)}$ and eigenvalues $\Lambda_\alpha$ are calculated by solving the eigenvalue problem  $\sum_{j=1}^\Omega \Delta_{i,j} v_i{(\alpha)} = \Lambda_\alpha v_i^{(\alpha)}$ with $\alpha=1,\ldots, \Omega$. The computed eigenvalues are real and non-positive. The eigenvectors are orthonormalized so to match the condition $\sum_{i}\sum_{\alpha,\beta} v_i^{(\alpha)} v_i^{(\beta)} = \delta_{\alpha, \beta}$.}. 

By inserting Eqs. (\ref{eq:modes}) into the linearized differential equation for the perturbations $\delta \phi_i$ and 
$\delta \psi_i$, one obtains the usual characteristic equation for $\lambda_{\alpha}$, which can be here cast in the form:  
\begin{equation}\label{eq:dispersion}
\begin{footnotesize}
\det\left( \begin{array}{cc}
f_\phi +\mu\left(1-\psi^*\right)\Lambda_\alpha -\lambda_\alpha & f_\psi +\mu\phi^*\Lambda_\alpha  \\
 & \\
g_\phi +\delta\psi^*\Lambda_\alpha & g_\psi +\delta\left(1-\phi^*\right)\Lambda_\alpha-\lambda_\alpha  \end{array} \right)=0
\end{footnotesize}
\end{equation}
where $f_q=\partial f/\partial q$ and $g_q=\partial g/\partial q$ for $q=\phi, \psi$. 

The Turing instability occurs, and the perturbation gets thus amplified, if $\lambda_{\alpha}(\Lambda_\alpha)$ is positive for some value of $\Lambda_\alpha$. In this respect, and as already remarked in \cite{nakao},  $\Lambda_\alpha$ plays the role of $-k^2$ for continuous media, where $k$ stands for the wavenumber of the plane wave mode. In Fig. \ref{fig:dispersion}(b), the dispersion relation is plotted for two distinct choices of the parameters (see legend). Symbols refer to the discrete linear growth rates $\lambda_\alpha$, as function of the corresponding Laplacian eigenvalues $\Lambda_\alpha$. The solid line represents the homologous dispersion relations, as obtained working within the continuous representation ($\Lambda_\alpha \rightarrow - k^2$). The upper curve (panel (b) of Fig. \ref{fig:dispersion}, circles) signals the presence of an instability. A significant fraction of the discrete rates $\lambda_{\alpha}$ is in fact positive. Conversely, the other profile (diamonds) is obtained for a choice of the chemical parameters that yields linear stability. By tuning the parameters, and evaluating the corresponding dispersion relation, one can eventually single out in a reference parameter space the region deputed to the instability. This is done in Fig.  \ref{fig:dispersion}(a) working in the plan $(b,c)$: the region of Turing instability, as predicted by the deterministic analysis, is filled with a uniform colour (yellow online). The (blue online) diamond falls outside the region of Turing order and points to the parameters employed in depicting the stable dispersion curve in panel (b). Similarly, the circle (magenta online) refers to the unstable profile. In this latter case, performing a direct integration of the mean-field equations (\ref{eq:mf}) one observes the spontaneous differentiation in activator-rich and activator-poor groups, as discussed in \cite{nakao}. A stochastic simulation can be also carried out, using an {\it ad hoc} implementation of the Gillespie Monte Carlo scheme \cite{gillespie}. In the movies annexed as supplementary material, the two dynamics, deterministic vs.  stochastic, are compared. Finite size fluctuations materialize in a modest perturbation ($\propto 1/\sqrt{N}$) of the idealized mean-field dynamics.  

Substantially different, is instead the scenario that is eventually recovered when comparing the simulations outside the region deputed to Turing instability. Setting the parameters ($b=76$,   $c=1076$) to the values ($b=76$, $c=1060$) that correspond to the diamond (blue online) of Fig. \ref{fig:dispersion}(a), the deterministic simulations always converge to the homogeneous fixed point, the concentrations of the species being therefore identical on each node of the network. At variance, a fragmentation into distinct groups is clearly observed in the stochastic simulations. The late time evolution of the stochastic system, as compared to the corresponding (trivial) deterministic solution, is displayed in Fig. \ref{fig:stocSim}. The effect of the stochastic driven polarization can be further realized when inspecting the annexed movies, which enables one to appreciate the full time evolution of the discrete dynamics. As for the case of continuous media, the endogenous stochastic noise is amplified and drives the formation of spatially extended, self-organized patterns outside the region of classical Turing order. Following \cite{biancalani}, we call these self-organized, asymptotically stable configurations, {\it Stochastic Turing patterns} on a network.

To gain analytic insight into the above mechanism, one can return to the van Kampen perturbative analysis and consider the next to leading approximation. One obtains a system of Langevin equations \cite{mckanePRL, biancalani} for the fluctuations $\xi_{si}$, $s=1,2$: 
\begin{equation}\label{eq:langevin}
\frac{d\xi_{si}}{d\tau}=\sum_{rj}  \mathcal{M}_{sr,ij} \xi_{rj}+\eta_{si}(\tau)
\end{equation}
where $\eta_{si}$ is a Gaussian noise with zero mean and correlator given by $\langle \eta_{s,i}(\tau)\eta_{rj}(\tau')\rangle=\mathcal{B}_{sr,ij}\delta_{\tau\tau'}$. The explicit form of the matrices $\mathcal{M}$
and $\mathcal{B}$ will be given in the Appendix. Define then following transformation:
  \begin{equation}\label{eq:trans}
\tilde{\xi}_\alpha=\sum_{i,\tau}\xi_i e^{-\mathtt{j}\omega\tau}v_i^{(\alpha)}
\end{equation} 
$\mathtt{j}$ denoting here the imaginary unit. The above operation is inspired to the Fourier transform: instead of 
expanding on a basis of plane waves, it is here natural to project the fluctuations along the $\Omega$ independent directions represented by the eigenvectors $v^{(\alpha)}$ of the discrete network Laplacian. One can hence define  a power spectrum of fluctuations of species $s=1,2$,  $P_s(\omega,\Lambda_\alpha)=\langle |\tilde{\xi}_s|^2\rangle$, in completely analogy with what it is customarily done 
in conventional Fourier analysis. In practical terms, the generalized power spectrum $P_s(\omega,\Lambda_\alpha)$ quantifies the portion of the signal power that is associated to given time ($\omega$) or spatial frequencies 
($\Lambda_\alpha$) range.  Some details of the calculations are given in the Appendix.
%%%%%%%%%%%%%%%%%%%%%%%%%%%%%%%%%%%%%%%%%%%%%%
\begin{figure}
\includegraphics[scale=0.3]{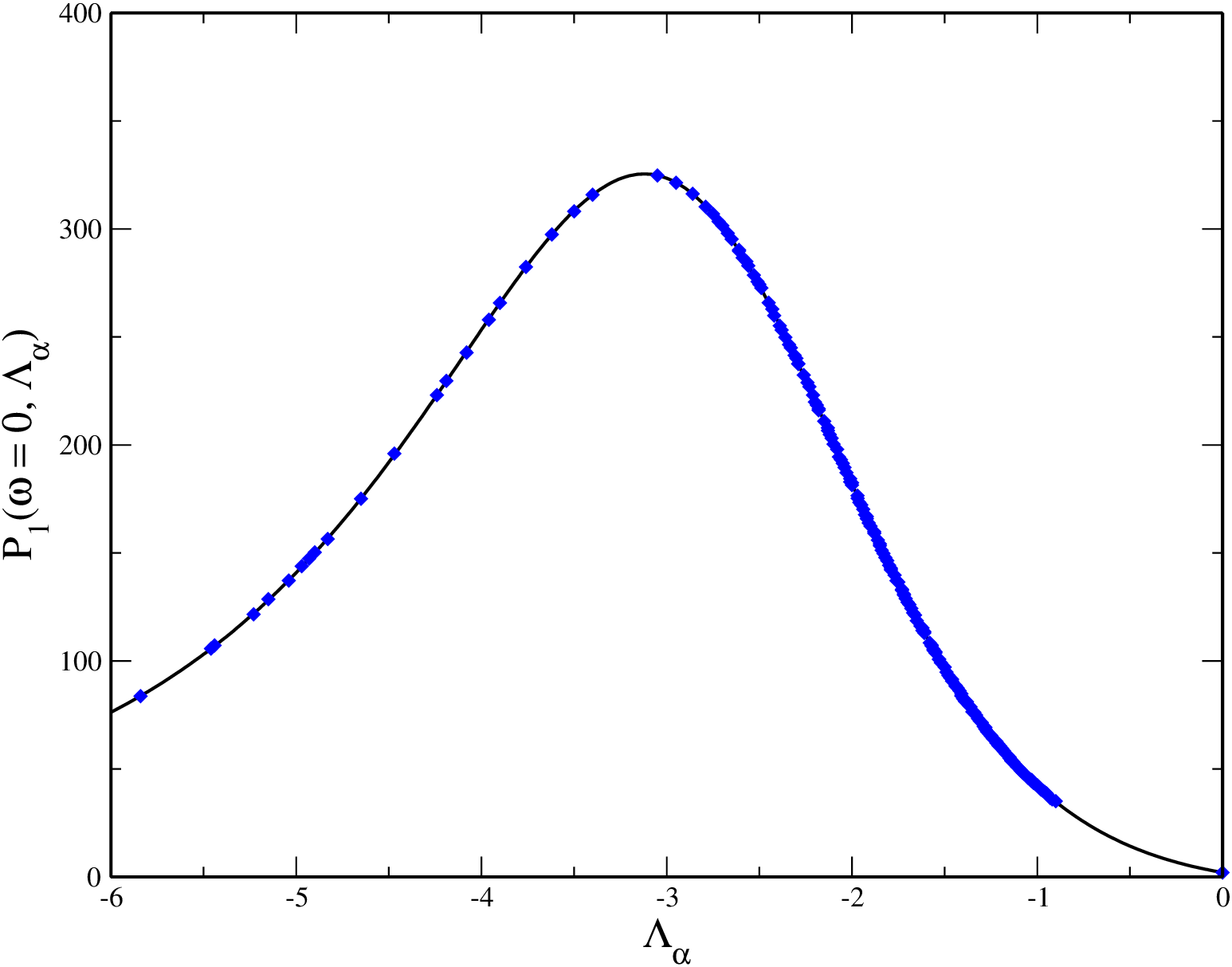} 
\caption{ Power Spectrum of fluctuations for species $X$  as a function of $\Lambda_\alpha$, $\omega=0$ (symbols). The solid line is the power spectrum calculated for a continuum media, i.e. when $\Lambda_\alpha$ is replaced by $-k^2$, where $k$ denotes the wavenumber of the plane wave mode. The curve refers to $a=d=1$, $b=76$, $c=1060$, $\mu=1$, $\delta=15$, a choice of parameters that correspond to operate outside the region of Turing instability (diamond in \ref{fig:dispersion}(a)). The network is constructed as specified in the caption of Fig. \ref{fig:dispersion}. 
\label{fig:PS}}
\end{figure}
%%%%%%%%%%%%%%%%%%%%%%%%%%%%%%%%%%%%%%%%%%%%%%

In Fig. \ref{fig:PS} the analytical power spectrum of species $X$, is plotted in the plane $\omega=0$, as a function of $\Lambda_\alpha$, for the parameters selection that corresponds to the diamond in Fig. (\ref{fig:dispersion}), i.e. outside the region of deterministic Turing instability. Symbols are obtained by sampling the  power spectrum over the discrete Laplacian eigenvalues $\Lambda_\alpha$. The solid line stands for the power spectrum calculated in the continuous limit when the discrete  $\Lambda_\alpha$ is replaced with its real counterpart $\rightarrow -k^2$ (see Appendix for a  discussion related to this point). A clear peak is displayed \footnote{Similar conclusions hold for species $Y$, the power spectrum.}, a finding that explains in turn the outcome of the stochastic based simulations reported in Fig. \ref{fig:stocSim}, proving on a formal ground that stochastic Turing patterns do exist on a network topology. 

In conclusion we have considered in this Letter the stochastic dynamics of the Brusselator model on a network. The model, representative of a broad class of systems that display Turing order in the mean field limit, has been investigated both analytically and numerically. In particular, we provide evidences on the intrinsic ability of the system to develop spatially heterogeneous configurations, outside the region of parameters classically deputed to Turing (deterministic) order. These self-organized patterns, reminiscent of the Turing instability, result from the spontaneous amplification of the demographic noise, stemming from the intimate discreteness of the scrutinized medium. Our analysis extends therefore the concept of Stochastic Turing order to the vast domain of network science, a discipline of paramount importance and cross-disciplinary interests. Further investigations can be planned working along these lines, and so explore the surprising degree of macroscopic order that can eventually originate from the noisy microscopic dynamics, for stochastic based systems defined on a network topology. As an example, stochastic travelling waves can be imagined to occur as a natural extension of \cite{biancalani2}. Incidentally, we also emphasize that the finite carrying capacity mechanism here imposed at the node level could turn useful to model those phenomena where jamming on a network topology are expected occur. 
\appendix
\section{}
At the next-to-leading order approximation in the van Kampen expansion, one obtains a Fokker Planck equation for the probability distribution of fluctuations, which is equivalent to the Langevin equation (\ref{eq:langevin}). For each fixed pair of nodes, $i$, $j$, the $2 \times 2$ matrix $\mathbf{\mathcal{M}_{sr,ij}}$  may be decomposed as the sum of two contributions, one relative to the activator-inhibitor reactions (non spatial components (NS)), and the other associated to the diffusion process (spatial components (SP)): $\mathbf{\mathcal{M}_{sr,ij}}=\mathbf{\mathcal{M}_{sr}}^{(NS)}+\mathbf{\mathcal{M}_{sr}}^{(SP)}
\Delta_{ij}$. The above matrices are  
evaluated at the mean-field fixed points $\phi^{*}$, $\psi^{*}$. 

After a lengthy calculation \cite{biancalani} one obtains the following entries for matrix $\mathbf{\mathcal{M}}^{(NS)}$: 
\begin{eqnarray}
\mathcal{M}^{(NS)}_{11}&=&-a-b-d+2c\phi^*\psi^*, \nonumber \\
\mathcal{M}^{(NS)}_{12}&=&-a+c\phi^{*^2}, \nonumber \\
\mathcal{M}^{(NS)}_{21}&=&b-2c\phi^*\psi^*, \nonumber \\
\mathcal{M}^{(NS)}_{22}&=&-c\phi^{*^2}. \nonumber 
\end{eqnarray}

The elements of matrix $\mathbf{\mathcal{M}}^{(SP)}$ read instead:  
\begin{eqnarray*}
\mathcal{M}^{(SP)}_{11}&=&2\mu(1-\psi^*), \\ 
\mathcal{M}^{(SP)}_{12}&=&2\mu\phi^*,   \\
\mathcal{M}^{(SP)}_{21}&=&2\delta\psi^*, \\  
\mathcal{M}^{(SP)}_{22}&=&2\delta(1-\phi^*). 
\end{eqnarray*}

As concerns the matrix $\mathbf{\mathcal{B}}$, one finds:
 \begin{eqnarray*}
\mathcal{B}_{11,ii} &=& D_1 + \tilde{k}_i H_1, \\
\mathcal{B}_{12,ii}&=&\mathcal{B}_{21, ii}=C \\
\mathcal{B}_{22,ii}&=& D_2 + \tilde{k}_i H_2    \\
\mathcal{B}_{11, ij}&=&-\left(\frac{1}{k_i}+\frac{1}{k_j}\right) H_1  \\ 
\mathcal{B}_{12, ij}&=&\mathcal{B}_{21, ij}=0. \\ 
\mathcal{B}_{22, ij}&=&-\left(\frac{1}{k_i}+\frac{1}{k_j}\right) H_2  
\end{eqnarray*} 
where: 
\begin{eqnarray*}
D_1 &=& a(1-\phi^*-\psi^*)+\phi^*\left(b+c\phi^*\psi^*+d\right)  \\
H_1 &=&  4\mu\phi^*(1-\phi^*-\psi^*)  \nonumber \\
C &=& -\phi^*\left(b+c\phi^*\psi^*\right) \nonumber \\
D_2 &=&  \phi^*\left(b+c\phi^*\psi^*\right) \nonumber \\
H_2 &=&  4\delta\psi^*(1-\phi^*-\psi^*) \nonumber \\
\end{eqnarray*}
Matrix $\mathcal{B}$ can be hence cast in the equivalent form:  
\begin{eqnarray}
\mathcal{B}_{ss,ij} &=& (D_s+\tilde{k}_i H_s) \delta_{ij} - \left(\frac{1}{k_i}+\frac{1}{k_j}\right) W_{ij} H_s \nonumber\\
\mathcal{B}_{rs,ij} &=& \mathcal{B}_{sr,ij} = C \delta_{ij}
\label{compact_for_B}
\end{eqnarray}
for $r,s=1,2$. 

Performing the transformation (\ref{eq:trans}) on both sides of the Langevin equation (\ref{eq:langevin}), we get  
\begin{equation}
 \mathtt{j} \omega\tilde{\xi}_{s}^\alpha = \sum_{r=1}^2\left(\mathcal{M}_{sr}^{(NS)}+\mathcal{M}_{sr}^{(SP)}\Lambda_\alpha\right)\tilde{\xi}_r^\alpha+\tilde{\eta}_s^\alpha
\end{equation} 
where the term $\tilde{\eta}_s^\alpha$ denotes the transform of the noise. 
Introducing the matrix 
\begin{equation}
\Phi_{sr}=\mathtt{j} \omega\delta_{sr}-\left(\mathcal{M}_{sr}^{(NS)}+\mathcal{M}_{sr}^{(SP)}\Lambda_\alpha\right)
\end{equation}
we get $\tilde{\xi}_s^\alpha =\sum_{r=1}^2 \Phi_{sr}^{-1}\tilde{\eta}_r^\alpha$, and thus the power spectrum of the $s$-th species is given by 
\begin{equation}
P_s(\omega,\Lambda_\alpha)=\langle |\tilde{\xi}^\alpha_s|^2\rangle=
 \sum_{r,k=1}^2 \Phi_{sr}^{-1}
\langle\tilde{\eta}^\alpha_r \tilde{\eta}_k^{\alpha}\rangle (\Phi_{ks}^{\dag})^{-1}
\end{equation}

It can be shown that $\langle\tilde{\eta}^\alpha_r  \tilde{\eta}_k^{\alpha} \rangle = \sum_{i,j}\mathcal{B}_{rk,ij}v_i^{(\alpha)}v_{j}^{(\alpha)} $.  An explicit form for the $2\times 2$ matrix $\langle\tilde{\eta}^\alpha_r  \tilde{\eta}_k^{\alpha} \rangle$ can be derived by making use of Eqs (\ref{compact_for_B}). Let us focus on the non trivial contribution $r=k$:
\begin{eqnarray*}
&&\sum_{i,j}\mathcal{B}_{rr,ij}v_i^{(\alpha)}v_{j}^{(\alpha)} =\\ &&\sum_{i,j} 
\left[ (D_r+\tilde{k}_i H_r) \delta_{ij} - \left(\frac{1}{k_i}+\frac{1}{k_j}\right) W_{ij} H_r \right] 
v_i^{(\alpha)}v_{j}^{(\alpha)} =\\ 
&& D_r \sum_{i} v_i^{(\alpha)} v_{i}^{(\alpha)} - H_r \sum_{i,j}  \Delta_{i,j} v_i^{(\alpha)} v_{i}^{(\alpha)} =
D_r - H_r \Lambda_{\alpha} 
\end{eqnarray*}
where in the last step we made use of $\sum_{i} v_i^{(\alpha)} v_{i}^{(\alpha)}=1$ and $\sum_j \Delta_{ij} v_j^{(\alpha)}=\Lambda_\alpha v_i^{\alpha}$. The component $r \ne k$ is trivially equal to $C$. 

The  power spectrum is fully specified as function of $\Lambda_\alpha$ and $\omega$. Figure \ref{fig:PS} is obtained by setting $\omega=0$ in the above formulae.  We emphasize that   the same result can be recovered from the continuum medium  power spectrum \cite{biancalani} provided $-k^2$ is replaced by the discrete eigenvalue $\Lambda_{\alpha}$.

We end this Appendix by providing a list of explanatory captions to
the movies annexed as supplementary material.
\begin{enumerate}
\item {\it mf\_inside\_X.mov} and {\it mf\_inside\_Y.mov} show respectively
the time evolution of the mean field concentrations $\phi_i$ and $\psi_i$.
The data are obtained by integrating the governing mean field
equations (\ref{eq:mf}).
Parameters are chosen so to yield  a Turing instability (magenta
circle in Fig.  \ref{fig:dispersion}).
At time $t=0$ a small perturbation is applied to perturb the
homogeneous fixed point. Then, the system evolves
toward a stable, non-homogeneous stationary configuration.
\item {\it st\_inside\_X.mov} and {\it st\_inside\_Y.mov} show the result
of the stochastic simulations
(fluctuating blue circles), for respectively $n_i/N$ and $m_i/N$. The
red symbols refer to a late time snapshot of the deterministic
dynamics. Parameters are set as specified above: the system is hence
inside the region of Turing instability. Notice that the noise that
takes the system away from the homogeneous fixed point is now
endogenous to the system and not externally imposed.
\item {\it st\_outside\_X.mov} and {\it st\_outside\_Y.mov} report on the
stochastic simulations (blue symbols), for respectively $n_i/N$ and
$m_i/N$. The parameters are now assigned so to fall outside the region
of Turing order (blue diamond in Fig. \ref{fig:dispersion}).
The mean field solutions are not destabilized and converge to the
stable fixed point. At variance, the stochastic simulations evolve
toward a non homogeneous state.

\end{enumerate}
\acknowledgements
Financial support from Ente Cassa di Risparmio di Firenze and the Program Prin2009 funded by Italian 
MIUR is acknowledged. D.F. thanks Tommaso Biancalani for pointing out reference \cite{nakao} and for stimulating discussions. F.D.P. thanks Alessio Cardillo for providing the code to generate the network.
\bibliographystyle{apsrev4-1}
\bibliography{bibliography}
\end{document}